\documentstyle[epsf]{mn}
 \input epsf.sty
\newif\ifAMStwofonts
\ifoldfss

  \ifCUPmtlplainloaded \else
    \NewTextAlphabet{textbfit} {cmbxti10} {}
    \NewTextAlphabet{textbfss} {cmssbx10} {}
    \NewMathAlphabet{mathbfit} {cmbxti10} {} 
    \NewMathAlphabet{mathbfss} {cmssbx10} {} 
  \fi
  \ifAMStwofonts
    \ifCUPmtlplainloaded \else
      \NewSymbolFont{upmath} {eurm10}
      \NewSymbolFont{AMSa} {msam10}
      \NewMathSymbol{\upi}     {0}{upmath}{19}
      \NewMathSymbol{\umu}     {0}{upmath}{16}
      \NewMathSymbol{\upartial}{0}{upmath}{40}
      \NewMathSymbol{\leqslant}{3}{AMSa}{36}
      \NewMathSymbol{\geqslant}{3}{AMSa}{3E}

      \let\leq=\leqslant 
       
    \fi
  \fi
\fi 

\ifnfssone
  \newmathalphabet{\mathit}
  \addtoversion{normal}{\mathit}{cmr}{m}{it}
  \addtoversion{bold}{\mathit}{cmr}{bx}{it}
  \newmathalphabet{\mathbfit} 
  \addtoversion{normal}{\mathbfit}{cmr}{bx}{it}
  \addtoversion{bold}{\mathbfit}{cmr}{bx}{it}
  \newmathalphabet{\mathbfss} 
  \addtoversion{normal}{\mathbfss}{cmss}{bx}{n}
  \addtoversion{bold}{\mathbfss}{cmss}{bx}{n}
  \ifAMStwofonts
    \ifCUPmtlplainloaded \else
      \UseAMStwoboldmath
      \makeatletter
      \new@mathgroup\upmath@group
      \define@mathgroup\mv@normal\upmath@group{eur}{m}{n}
      \define@mathgroup\mv@bold\upmath@group{eur}{b}{n}
      \edef\UPM{\hexnumber\upmath@group}
      \new@mathgroup\amsa@group
      \define@mathgroup\mv@normal\amsa@group{msa}{m}{n}
      \define@mathgroup\mv@bold\amsa@group{msa}{m}{n}
      \edef\AMSa{\hexnumber\amsa@group}
      \makeatother
      \mathchardef\upi="0\UPM19
      \mathchardef\umu="0\UPM16
      \mathchardef\upartial="0\UPM40
      \mathchardef\leqslant="3\AMSa36
      \mathchardef\geqslant="3\AMSa3E

      \let\leq=\leqslant 

    \fi
  \fi
\fi 

\ifnfsstwo
  \DeclareMathAlphabet{\mathbfit}{OT1}{cmr}{bx}{it}
  \SetMathAlphabet\mathbfit{bold}{OT1}{cmr}{bx}{it}
  \DeclareMathAlphabet{\mathbfss}{OT1}{cmss}{bx}{n}
  \SetMathAlphabet\mathbfss{bold}{OT1}{cmss}{bx}{n}
  \ifAMStwofonts
    \ifCUPmtlplainloaded \else
      \DeclareSymbolFont{UPM}{U}{eur}{m}{n}
      \SetSymbolFont{UPM}{bold}{U}{eur}{b}{n}
      \DeclareSymbolFont{AMSa}{U}{msa}{m}{n}
      \DeclareMathSymbol{\upi}{0}{UPM}{"19}
      \DeclareMathSymbol{\umu}{0}{UPM}{"16}
      \DeclareMathSymbol{\upartial}{0}{UPM}{"40}
      \DeclareMathSymbol{\leqslant}{3}{AMSa}{"36}
      \DeclareMathSymbol{\geqslant}{3}{AMSa}{"3E}

      \let\leq=\leqslant 

    \fi
  \fi
\fi 

\ifCUPmtlplainloaded \else
  \ifAMStwofonts \else 
    \def\upi{\pi}
    \def\umu{\mu}
    \def\upartial{\partial}
  \fi
\fi

\title[Permanent Superhumps in V1974~Cyg]
  {Permanent Superhumps in Nova V1974~Cygni~1992}

\author[A. Retter, E.M. Leibowitz and E.O. Ofek]
   {A.~Retter,$^1$
   E.M.~Leibowitz
   and E.O.~Ofek\\
  School of Physics and Astronomy and the Wise Observatory,
Raymond and Beverly Sackler Faculty of Exact Sciences,\\
Tel-Aviv University, Tel Aviv, 69978, Israel\\
   $^1$ email: alon@wise.tau.ac.il\\}

\date{submitted 1996 September 5}
\date{accepted 1996 November 4}
\pagerange{\pageref{firstpage}--\pageref{lastpage}}
\pubyear{1997}

\def\LaTeX{L\kern-.36em\raise.3ex\hbox{a}\kern-.15em
    T\kern-.1667em\lower.7ex\hbox{E}\kern-.125emX}

\begin{document}

\label{firstpage}

\maketitle

\begin{abstract}
 We present results of 32 nights of CCD photometry of V1974~Cygni, from the
years 1994 and 1995. We verify the presence of two distinct periodicities in
the light curve: 0.0812585 day $ \approx 1.95$ hours and 0.0849767 d $
\approx 2.04$ hr. We establish that the shorter periodicity is the orbital
period of the underlying binary system. The longer period 
oscillates with an average value of
$|\dot{P}|\sim 3\times 10^{-7}$---typical to permanent superhumps.

The two periods obey the linear relation between the orbital and superhump
periods that holds among members of the SU Ursae Majoris class of dwarf novae.
A third periodicity of 0.083204 d $\approx 2.00$ hr appeared in 1994 but not
in 1995. It may be related to the recently discovered anti-superhump
phenomenon. These results suggest a linkage between the classical nova 
V1974~Cyg and the SU UMa stars, and indicate the existence of an accretion disk
and permanent superhumps in the system no later than 30 months after the 
nova outburst. 

From the precessing disk model of the superhump phenomenon we estimate that 
the mass ratio in the binary system is between 2.2 and 3.6. Combined with 
previous results this implies a white dwarf mass of 0.75--1.07 M$_{\odot}$.

 \end{abstract}

 \begin{keywords}

 novae - stars: individuals: V1974~Cygni - accretion disks - stars: binaries:
 eclipsing - stars: oscillations - stars: white dwarf

 \end{keywords}

 \section{Introduction}

 \subsection{V1974~Cygni}

 V1974~Cygni was discovered on February 19 1992 by Collins (1992). It was the
 brightest nova since V1500~Cygni, and was soon recognized as a fast one, with
 $t_{2V}\approx 16d$, and $t_{3V}\approx 40d$. The prenova was estimated as
 $m_{V}\approx 19.5$, indicating an outburst amplitude of about 15 magnitude
 (Annuk, Kolka \& Leedjarv 1993). A massive multiwavelengths study of the star
 has been carried out in the last 5 years, including quite a few optical
 photometric observations (Hurst 1992, Kidger 1992, Chochol et al. 1993, 
DeYoung \& Schmidt 1993, 1994, Semeniuk et al. 1994, 1995, 
Retter, Ofek \& Leibowitz 1995, Retter, Leibowitz \& Ofek 1996). 

 Periodic oscillations in the light curve (LC) of the star were discovered in
 1993 by DeYoung \& Schmidt (1993, 1994). They detected a period of
 0.081263$\pm$0.000003 day in the I band. The modulation was about
 0.16$\pm$0.05 mag, and its shape was not symmetric, as the rise to maximum was
 faster than the decline to minimum. The variation was much weaker in the V
 band (0.05$\pm$0.06 mag). They interpreted the periodicity as the orbital
 period, and the modulation as resulting from the varying aspect of the
 illuminated hemisphere of the companion by the intense radiation of the hot
 nova.

 Semeniuk et al. (1994) reported the detection of a different periodicity
 (0.0850 day) from observations in the V band during 1994. The amplitude of
 this variation was about 0.05 mag, and it had a similar amplitude in the I and
 the R bands. They also noticed that this period was not stable.

 The two distinct periodicities in the LC of V1974~Cyg were also independently 
discovered by Retter et al (1995). They suggested a connection between
 the second periodicity of the nova and the superhump (SH) phenomenon in the SU
 UMa stars.

 In a second publication, Semeniuk et al. (1995) confirmed the continued
 presence of two distinct periodicities in the LC of the nova, the longer of
 which was decreasing in time. They concluded that the shorter, stable period
 is the orbital period of the nova system, and attributed the longer one to the
 spin period of white dwarf (WD) in the system. In their model, the WD
 possesses a strong magnetic dipole field, the axis of which is inclined with
 respect to the spin axis of the star. The nova outburst caused the spin of the
 WD to get out of synchronization with the orbital revolution. The light
 variation of the longer period results from a distinct light source (an
 accretion column) near one of the magnetic poles of the star. The spin period
 of the WD evolves back into synchronization, hence the change in the longer
 periodicity.

 Retter et al (1996) suggested instead, that the longer photometric
 period of V1974 Cyg is a disk feature, related to the well known superhump
phenomenon in SU UMa stars. In this work we present further evidence in 
support of our interpretation of the longer period of V1974~Cyg as a SH 
phenomenon.

 \subsection{The superhump phenomenon in novae and in other CVs}

 The SH phenomenon is the appearance of a periodic or a quasi-periodic
 modulation in the LC of the SU UMa class of dwarf novae, during the
 superoutburst events that characterize the photometric behaviour of these
 stars. The period of the SH is a few percent longer than the
 orbital period of the system in which it is observed (laDous 1993).
 Stolz \& Schoembs (1981, 1984) found a linear relation between
 the relative excess of the SH period over the orbital one, and the orbital
 period. The commonly accepted interpretation of the phenomenon is that the
 light modulations result from the precession of the accretion disc around the
 WD of the underlying binary system. The observed SH periodicity is in fact the
 beat of the disc precession period in the co-rotating frame and the orbital
 period of the binary system (Whitehurst 1988, Osaki 1996).

 Recently it was suggested that a few classical novae show SH-like
 characteristics in their optical LCs. For example, Patterson \& Richman (1991),
 Patterson et al. (1993c) and Thomas (1993) suggested a SH interpretation 
for the double
 periodicity that was observed in the LC of the classical nova V603 Aquilla. A
 similar interpretation was given to a similar phenomenon in the LC of nova CP
 Puppis (White \&  Honeycutt 1992, White, Honeycutt \& Horne 1993 and 
Thomas 1993). Patterson et al. (1993c) suggest also a few other classical 
novae as potential superhumpers.

 The SH phenomenon was also invoked as an interpretation of a few peculiarities
 in the LC of AM CVn's stars (Patterson, Halpern \& Shambrook 1993b, 
Patterson et al. 1993c
 and Solheim et al. 1996) and in the photometric time series of a few soft
 x-ray sources (White 1989, Charles et al. 1991, Bailyn 1992, Zhang \& Chen
 1992 and Kato, Mineshige \& Hirata 1995).

 Unlike the case in the SU UMa stars, where the SH modulation is a transient
 feature, in a few of the other CV stars, such as V603 Aql, CP Pup and AM CVn,
 the systems are believed to experience permanent SHs. The existence of
 this mode of variation has been theoretically established by Osaki (1996).

 \section{Observations}

 We observed V1974~Cygni during 32 nights from September 1994 to November 1995.
 Table 1 presents a summary of the observations schedule. The photometry was
 carried out with the 1 meter telescope at the Wise Observatory, using the
 Techtronix 1K CCD camera, described in Kaspi et al. (1995). 
During the first
 three nights we switched successively between the standard V,R and I filters.
 In the fourth night the photometry was in the B band, and from then on we used
 only the I filter. Our I filter is somewhat redder than the standard bandpass,
 because its red end is determined by our CCD camera cut-off, which has a
 longer tail into lower frequencies. The typical exposures times were 
between 1 to 3 minutes. The number of frames obtained in each filter on 
our programme is: 2620 (I), 100 (R), 96 (V) and 67 (B).

\begin{table}
  \caption{The Observations Time Table}
  \begin{tabular}{@{}cccc@{}}

UT&Time of Start&Run Time&Points\\
   Date  &(JD+2449000) & (hours)&number\\
\\

240994& 620.719&      6.5 &144\\
251094& 651.699&      5 &140\\
011194& 658.682&      0.25 &3\\
081194& 665.656&      6 &168\\
061294& 693.652&      5.5 &126\\
071294& 694.649&      5.5 &124\\
081294& 695.656&      5 &93\\
091294& 696.669&      5 &90\\
101294& 697.672&      5  &90\\
240595& 862.907&      4.5 &65\\
250595& 863.892&      4.5 &84\\
060695& 875.967&      2.5 &41\\
080695& 877.843&      5.5 &100\\
090695& 878.831&      6 &109\\
100695& 879.829&      6 &96\\
110695& 880.823&      6 &114\\
250795& 924.778&      7.5 &135\\
260795& 925.780&      7.5 &140\\
270795& 926.780&      7.5 &146\\
150995& 976.725&      7 &160\\
160995& 977.703&      7 &162\\
011095& 992.697&      1 &38\\
021095& 993.699&      1 &26\\
141095& 1005.679&      5 &93\\
181095& 1009.718&      3.75 &69\\
191095& 1010.682&      3.25 &62\\
201095& 1011.685&      3.5 &38\\
211095& 1012.733&      5 &88\\
161195& 1038.663&      5.5 &98\\
171195& 1039.644&      1.5 &28\\
181195& 1040.655&      1 &18\\
191195& 1041.662&      5 &95\\

\end{tabular}
\end{table}

 Photometric measurements were performed using the DAOPHOT program (Stetson
 1987). Instrumental magnitude of the nova, as well as of 3 to 20 reference
 stars, depending on the site photometric conditions, were obtained from each
 frame. An internal consistent series of nova magnitudes was obtained by using
 the Wise Observatory reduction program DAOSTAT (Netzer et al. 1996).

 Fig. 1 displays a comprehensive LC of the nova in the V band from
 outburst to December 1995. The data was taken from AFOEV database, CDS,
 France. The lower marks in the figure indicate the times of our 
observations. They
 are divided into three sub-groups: September--December, 1994 (R1),
May--July, 1995 (R2), and September--November, 1995 (R3). The times of
 observations by DeYoung \& Schmidt (1994), and by Semeniuk et al. (1995) are
 marked in the figure as well.

\begin{figure}

\centerline{\epsfxsize=3.5in\epsfbox{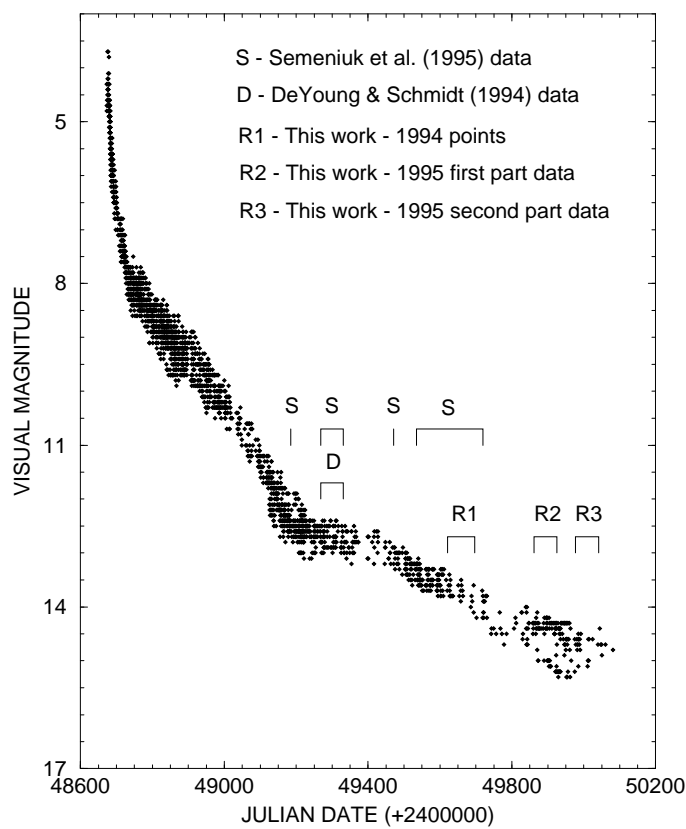}}

 \caption{The light curve of Nova~Cyg~92. Data points are visual estimates of
 amateur astronomers, compiled by AFOEV. The times of observations by DeYoung
 \& Schmidt (1994), and by Semeniuk et al. (1995) are marked, along with the
 times of our own observations. The latter are divided into three sub-groups:
 September--December, 1994 (R1), May--July, 1995 (R2), and 
September--November, 1995 (R3).}
 \end{figure}

 Figure 2 presents the entire I LC of the nova, as measured in our observing
 programme. One can see that during the time interval spanned by 
our observations, the
 nova declined by about 0.5 magnitude. There is also some variation from night
 to night in the nightly average magnitude. Some of it may be related to the
 beat period between the two periods of $\sim2$ hours that characterize the LC
 of the nova (see below).

\begin{figure}

\centerline{\epsfxsize=3.5in\epsfbox{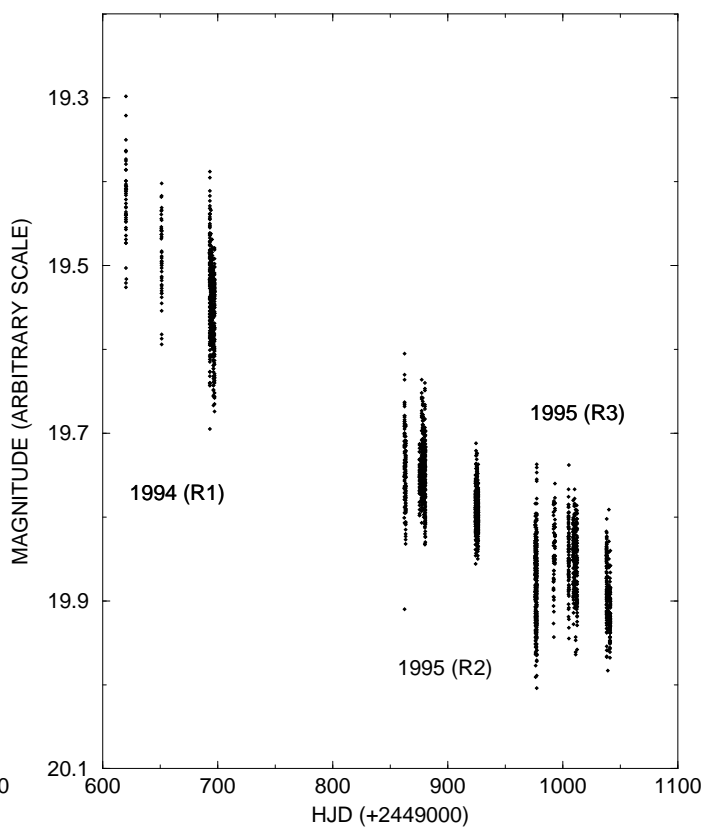}}

 \caption{All I magnitudes of V1974~Cyg measured in this work, divided into
 three sub-groups.}
 \end{figure}

 The LC of the nova in all four passbands exhibits a sinusoidal like
 modulation. Although our observations in the B,V and R bands are much
 fewer than in I, we still can appreciate that the peak to peak amplitude of
 the variation seems to be similar in all bands, varying between 0.05 and 0.10
 magnitude from night to night. Thus it appears that the short term optical
 light variations, of about 2 hours time scale, are independent of colour.
 Closer examination of the modulation reveals that it is asymmetric, with a
 rise to maximum that is slower than the decline to minimum. We note that this
 structure is different from the one observed previously by DeYoung \& Schmidt
 in 1993.

\begin{figure}

\centerline{\epsfxsize=2.5in\epsfbox{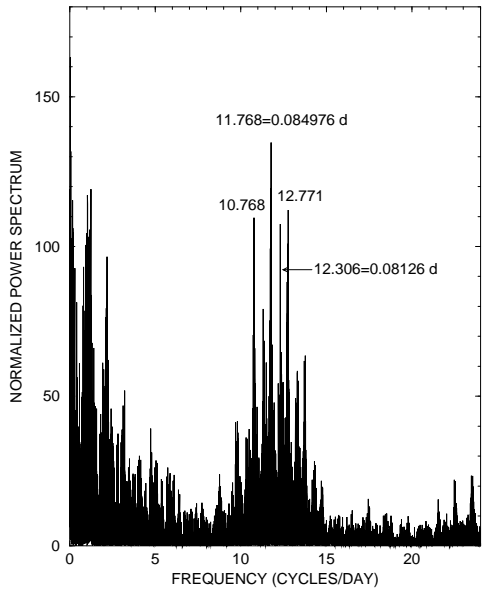}}

 \caption{Normalized power spectrum of the entire I LC. The two double
 structures of peaks around the central frequencies 11.768 and 12.306 is the
 pattern of 1, 1/2, 1/3 etc. day aliases of these periodicities.}
\end{figure}

 \section{Data Analysis}

 \subsection{The two major periodicities}

 Figure 3 is a plot of the normalized power spectrum (PS) (Scargle 1982) of all
 our 2620 I band points, after dereddening by subtracting a second degree
 polynomial from the data. The PS is dominated by two similar patterns of
 1,1/2,1/3 etc. day aliases around the two central frequencies of
 11.76792 day$^{-1}$ (P=0.0849767$\pm$0.0000005 day), and 12.30640 day$^{-1}$
 (P=0.0812585$\pm$0.0000005 day). See also Figures 4 and 7 for a magnification
 of the relevant range.

\begin{figure}

\centerline{\epsfxsize=3.5in\epsfbox{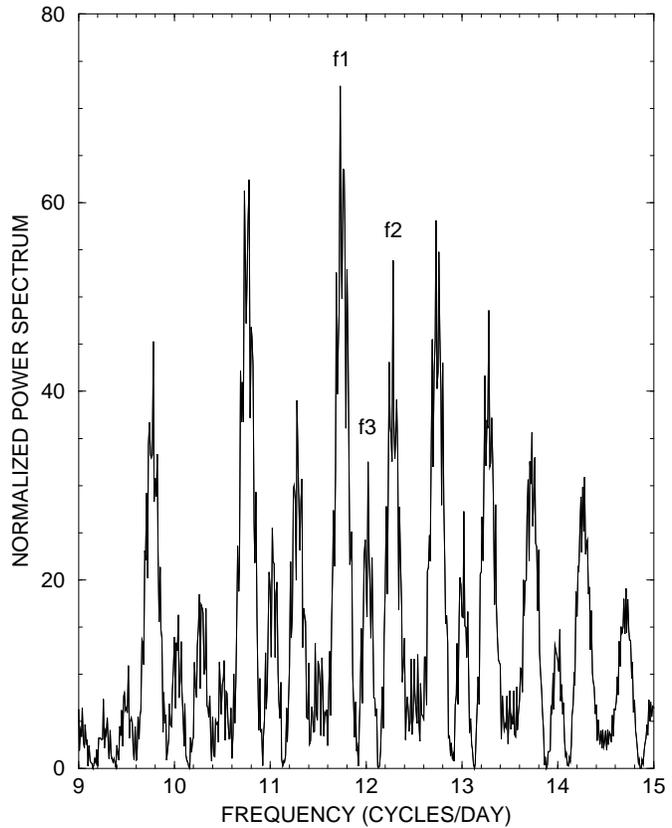}}

 \caption{Normalized power spectrum calculated from the I band points in 1994.
 Two similar patterns of peaks around the central frequencies 11.768 (f1) and
 12.306 (f2) are clearly seen, as in Fig. 3. A third lower peak around the
 frequency 12.019 (f3) indicates the presence of another true periodicity in the
 LC---see text.}
\end{figure}

The group of peaks at the lower end of the PS corresponds to periodicities,
that are longer than one night in our observing program. Their presence in the
PS is sensitive to the dereddening procedure and therefore their reality cannot 
be determined from our data.

 In order to check the reliability of the two independent periodicities in the 
LC of the nova, we subtracted from the data the first harmonic of the 0.0849767
 day period. The PS of the residuals shows clearly the shorter,
 0.0812585 day periodicity as a dominant peak in the PS. Conversely, when the
 shorter period is removed from the data, the longer one remains clearly in the
 residual LC. As a further check we created an artificial LC on the times of
 the real observations, by superposing a sine wave of one of the two
 periodicities over a random distribution of points representing white noise.
 The PS of each of these synthetic LCs showed only the corresponding planted
 periodicity, surrounded by an alias pattern similar to the one of the real
 data, with no trace of the other periodicity.

 \subsection{Other periodicities}

 We further checked the data for possible additional short term periodicities.
 This was done by computing the PS of the LC of the star, in each
 of the individual nights of our observations. These power spectra were then
 added together to form an average nightly PS of the star (see Skillman, 
Patterson \& Thorstensen 1995, Patterson 1995). No periodicity in the range 
from the Nyquist frequency to the longer period of about 2 hours stands out 
in this PS.

 We also examined the harmonics of the main 11.76792 frequency in a motivation
 to see whether they are displaced towards the blue, an effect, which is
 reported in a few SU UMa systems (Patterson et al. 1995a and Harvey \&
 Patterson 1995).  The second harmonic is found at 23.61 cycles/day---about 
0.3 percent larger than twice the fundamental frequency. The third harmonic 
is 35.36 cycles/day---0.2 percent
 positively misplaced. The higher harmonics did not rise above the noise level.
 Most of the harmonics of the three different parts of the data mentioned
 above showed similar shifts, but the effect was less obvious.

 \subsection{The 1994 light curve}

 Fig. 4 displays the normalized PS of the 1994 I data, zoomed on
 the close neighborhood of the two frequencies discussed in a previous
 section. The two peaks, corresponding to these frequencies, along with the
 daily alias pattern around each one of them, are clearly seen in the figure.
 However, a third periodicity of 0.083204$\pm$0.000001 day, together with its
 own one day alias pattern, is also present in the PS. This pattern of triple
 frequencies is absent from the PS of the 1995 data, in which only the two
 major periods stand out, as in Figure 3. We checked the independence of the
 third periodicity that appears in 1994, with the same two methods discussed
 above. We found that during the 1994 observational season, the LC of the
 nova has indeed been modulated by a third, independent periodicity of 0.083204
 day, in addition to the other two major periods. Due to the scarcity of the
 1994 data, however, and since the third period did not re-appear in 1995, we
 must regard this conclusion as uncertain.

 \subsection{Structure of the 2 periodicities}

 In Fig. 5 we show the I band data, folded onto the 0.0849767 day period. The
 points are the average magnitude value in each of 40 equal bins that cover the
 0-1 phase interval. The bars are the $1 \sigma$ uncertainties in the value of
 the average values. The peak to peak amplitude of the average variation is
 about 0.055$\pm$0.010 magnitude. The periodic photometric cycle is clearly
 asymmetric, with a slow, stepwise rise to maximum and a fast decline to
 minimum. Similar values for the variation amplitude are deduced from the 
different three parts of our data discussed above.

\begin{figure}

\centerline{\epsfxsize=3.5in\epsfbox{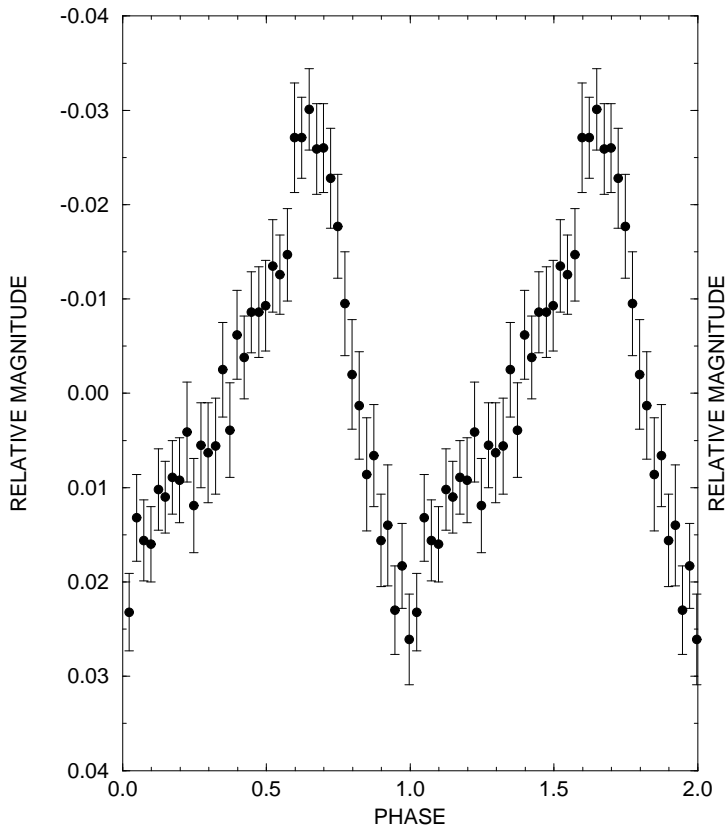}}

 \caption{The I filter light, folded onto the 0.0849767 day period, and
 binned into 40 equal bins.}
\end{figure}

 Fig. 6 displays the LC, folded onto the shorter 0.0812585 day period,
 with the same binning and symbols as in Figure 5. This LC is rather sinusoidal
 in shape, with a peak to peak amplitude of about 0.042$\pm$0.010 magnitude.
Again, within the error limits, the amplitude of the variation during the three
parts of the observations is the same.

 A noteworthy feature seen in Fig. 6 is the brief shallow dip in the light
 curve at phase 0.5. Its reliability was checked by subtracting a pure
 sinusoidal from the folded and binned data, resulted in a distribution of the
 points around the zero level, with three adjacent points lower than the others
 at the relevant phase. These three points were at 1.2, 2.4 and 1.5$\sigma$
from the expected value. The chances to find such points together around a
unique area at the LC (phase 0.5) are less than 0.01 percent.
All folded LCs of the three parts of the data (September--December,
 1994, May--July, 1995, and September--November, 1995) show similar behaviour
 at this phase.

Our data is consistent with the notion that there is no 
long-term variation in the amplitude of the two periodicities, 
derived from the three epochs of our observations. 
However, comparing these results with the values, determined by 
DeYoung \& Schmidt (1994) and by Semeniuk et al. (1995), it is clear that 
the amplitude of the photometric binary variation decreased sharply from 
1993 to 1994. 

\begin{figure}

\centerline{\epsfxsize=3.5in\epsfbox{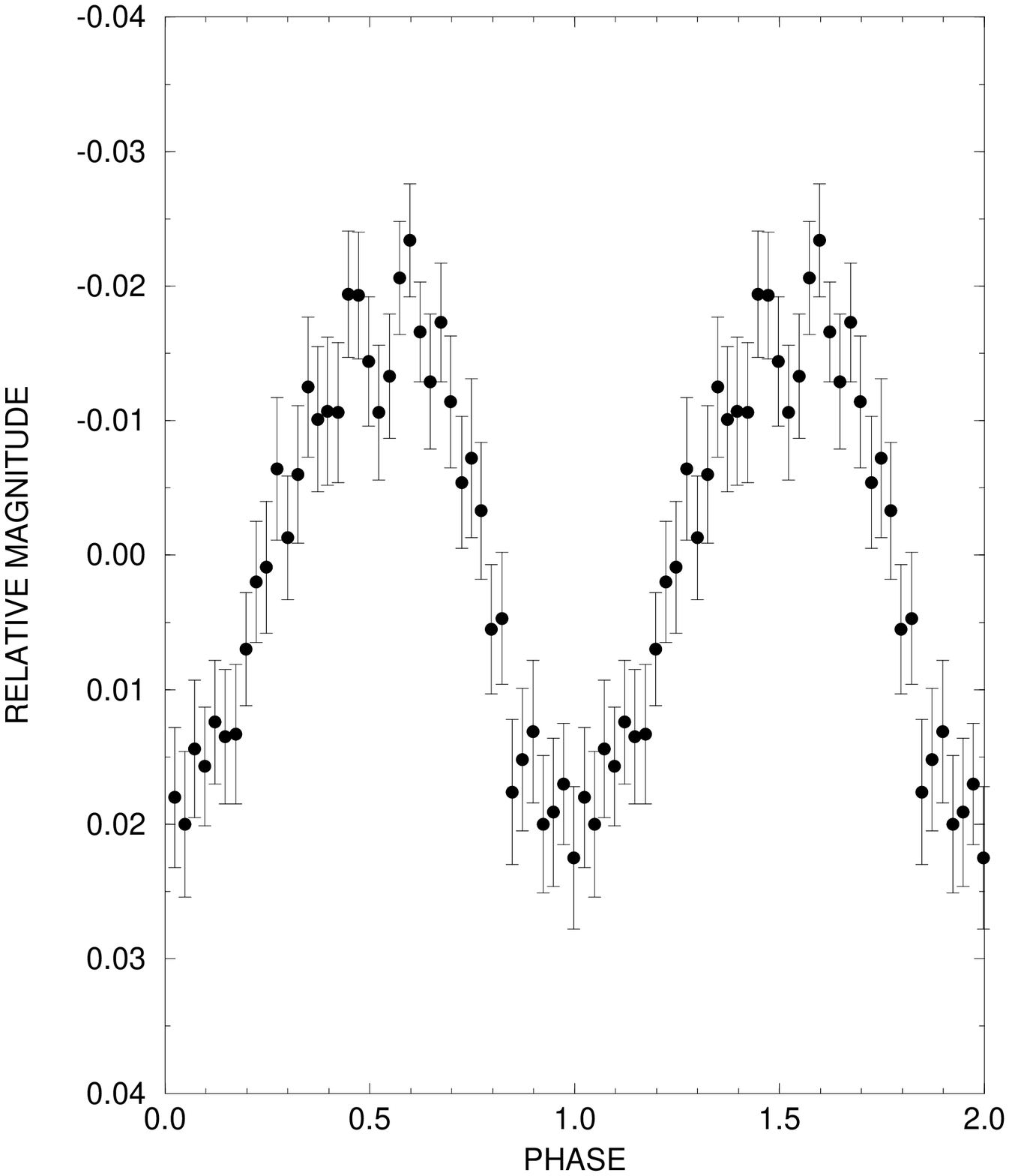}}

 \caption{The light curve, folded onto the shorter 0.0812585 day period.
Binning is as in Fig. 5.
Note the eclipse-like shape at phase 0.5---half a cycle after minimum
of light.}
\end{figure}

\subsection{Coherence of the two periodicities}

 Fig. 7 is a zoom picture of two sections of the PS of V1974~Cyg seen in
 Figures 3 \& 4, in the immediate neighborhood of the two major frequencies of
the system. It is quite noticeable that the higher frequency is represented 
by a distinct, single peak, while the lower frequency, although appearing with 
more power in the spectrum, is represented by a group of peaks rather than by
 one. The light variation corresponding to this frequency should therefore be
 rather regarded as a quasi-periodicity.

 In view of this difference we hereby give the best fitted ephemeris of the
 shorter periodicity:\\
 \\
 T${min}$ = HJD 2449693.2117 + 0.0812585 E.\\
 \hspace*{1.1in}$\pm $0.0005 \hspace*{0.01in}  $\pm$ 0.0000005 \\

 \begin{figure}

\centerline{\epsfxsize=2.5in\epsfbox{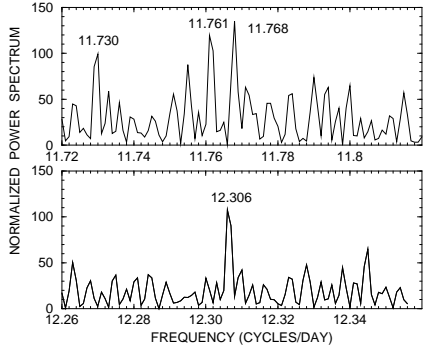}}

 \caption{The immediate vicinity of the two frequencies in the power
 spectrum of V1974~Cyg. The shorter 11.76792 frequency, seen at the upper
 panel, has several neighboring peaks with similar power in each. 
The higher 12.30640
 frequency at the lower panel is a single isolated peak in the graph.}
\end{figure}

 In order to characterize further the nature of the quasi-periodicity around
 the lower frequency 11.76792 day$^{-1}$, we plot in Figure 8 the O-C values of
 times of maximum in the LC of the nova. We derived times of maxima from our
 own observations by considering a third order polynomial that was fitted by
 least squares to the observed points around phases of maximum. A third degree
 was adopted because of the asymmetrical shape of the variation in the cycle
 (see Figure 5). Following Semeniuk et al. (1995), we took into account maxima
 from all filters. The maxima are listed in Table 2. The 1$\sigma$ errors are
 of the order of 0.005 day.

\begin{figure}

\centerline{\epsfxsize=3.5in\epsfbox{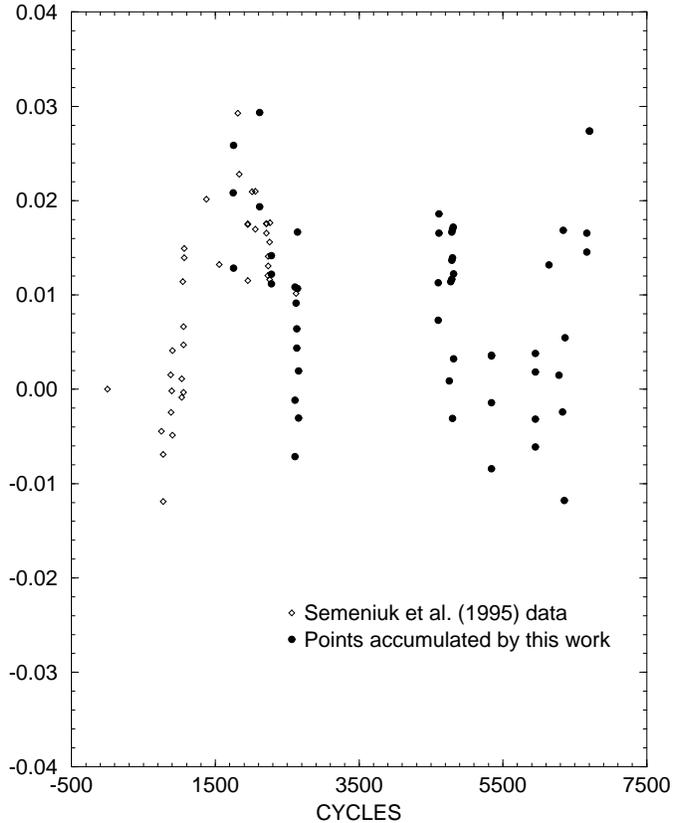}}

 \caption{O-C diagram of the 0.0849767 day period. Two obvious changes in
$\dot{P}$ are seen: a decrease of the period around cycle 2000 and an
increase at about cycle 6000.}
\end{figure}

 \begin{table}
  \caption{Maxima times of the 0.0849767 day period}
  \begin{tabular}{@{}lclr@{}}

    HJD     & Filter & Cycle & O - C  \\
            &        &       &(days) \\
            &        &       &   \\
2449620.27  &   I    &  1750 &  0.021 \\
2449620.347 & V,R,I  &  1751 &  0.013 \\
2449620.445 &  V,I   &  1752 &  0.026 \\
2449651.295 &  V,I   &  2115 &  0.029 \\
2449651.37  &   I    &  2116 &  0.019 \\
2449665.216 &   B    &  2279 &  0.014 \\
2449665.298 &   B    &  2280 &  0.011 \\
2449665.384 &   B    &  2281 &  0.012 \\
2449693.17  &   I    &  2608 &  0.011 \\
2449693.243 &   I    &  2609 & -0.001 \\
2449693.322 &   I    &  2610 &  0.007 \\
2449694.273 &   I    &  2621 &  0.009 \\
2449695.203 &   I    &  2632 &  0.010 \\
2449695.290 &   I    &  2633 &  0.006 \\
2449696.235 &   I    &  2644 &  0.017 \\
2449696.314 &   I    &  2645 &  0.011 \\
2449697.24  &   I    &  2656 &  0.002 \\
2449697.32  &   I    &  2657 & -0.003 \\
2449862.444 &   I    &  4600 &  0.007 \\
2449862.525 &   I    &  4601 &  0.007 \\
2449863.469 &   I    &  4612 &  0.017 \\
2449863.556 &   I    &  4613 &  0.019 \\
2449875.52  &   I    &  4754 &  0.001 \\
2449877.40  &   I    &  4776 &  0.011 \\
2449878.34  &   I    &  4787 &  0.017 \\
2449878.420 &   I    &  4788 &  0.012 \\
2449878.507 &   I    &  4789 &  0.014 \\
2449879.34  &   I    &  4799 &  0.003 \\
2449879.442 &   I    &  4800 &  0.014 \\
2449879.53  &   I    &  4801 &  0.017 \\
2449880.380 &   I    &  4811 &  0.017 \\
2449880.451 &   I    &  4812 &  0.003 \\
2449880.545 &   I    &  4813 &  0.012 \\
2449925.319 &   I    &  5340 &  0.004 \\
2449925.392 &   I    &  5341 & -0.008 \\
2449925.484 &   I    &  5342 & -0.001 \\
2449925.574 &   I    &  5343 &  0.004 \\
2449977.240 &   I    &  5951 &  0.004 \\
2449977.318 &   I    &  5952 & -0.003 \\
2449977.408 &   I    &  5953 &  0.002 \\
2449977.485 &   I    &  5954 &  0.006 \\
2449993.225 &   I    &  6139 &  0.013 \\
2450005.28  &   I    &  6281 &  0.002 \\
2450009.27  &   I    &  6328 & -0.002 \\
2450010.224 &   I    &  6339 &  0.017 \\
2450011.47  &   I    &  6354 & -0.012 \\
2450012.422 &   I    &  6365 &  0.006 \\
2450038.264 &   I    &  6669 &  0.015 \\
2450038.351 &   I    &  6670 &  0.017 \\
2450041.251 &   I    &  6704 &  0.027 \\
2450041.336 &   I    &  6705 &  0.027 \\

 \end{tabular}
 \end{table}

 We get similar results when we fit to the data three harmonics of the two
 periods simultaneously, and after subtracting the harmonics of the shorter
 period from the data. In this way the effect of the shorter periodicity is
 nearly eliminated. The maxima in the LC of the residuals are less
 distinguishable than the maxima identified directly in the raw data, but the
 O-C graph remains essentially unaltered.

 The plot in Fig. 8 includes in addition to our own maxima values also times of
 maxima determined by Semeniuk et al. (1995). Neglecting the first point, we
 can see at least two changes in the periodicity---a decrease of the period at
 the left hand side of the diagram in 1994, and an increase at the right hand
 side during 1995. Additional changes in the sign of $\dot{P}$ may have occurred
 at the gap in the observations, during the beginning of 1995.

 It should be mentioned that the O-C diagram of the minima shows very similar
 features. We also checked the increase of the period during 1995, by dividing
 the data of that year into two sub-groups (May--July, 1995, and 
September--November, 1995). 
We then fitted simultaneously three harmonics of the two
 periodicities to the data, and subtracted the terms of the higher, assumed
 fixed, 12.30640 frequency. The PS of the residuals of the first,
 earlier group yielded a period of 0.084955 day, while the peak of the PS of 
the late group was at about 0.085005 day. This check confirms the
 increase of the period during 1995, and fits exactly the periods values,
 derived from the O-C diagram, assuming constant $\dot{P}$ during the two
 epochs.

Fig. 8 demonstrates that the longer period oscillates around a certain value,
so its mean derivative over a long time should be zero. However, in order 
to give a general estimation of $\dot{P}$, we considered just the 
1995 data, in which the period increased. The result is:

$\dot{P} \sim (0.085005-0.084955) / 180 \sim 3 \times 10^{-7}$.\\
Similar absolute value with the opposite sign describes the period change 
in 1994.

\section{Discussion}

\subsection{The two periodicities in the LC of the nova}

 The large amount of continuous photometric data on V1974~Cygni, accumulated by
 DeYoung \& Schmidt (1994), Semeniuk et al. (1994, 1995), and in this work
 confirms undoubtedly the presence of two independent periods---0.0812585 day
 and 0.0849767 day in the LC of this nova. The persistence of the shorter
 period in the LC during more than two years, its presentation in the PS as an
 isolated narrow peak, and the structure of its average cycle, with the
 apparent dip at phase 0.5, suggest strongly that it is the orbital period of
 the binary system.

 While there is a general agreement that the shorter period is indeed the
 binary period of the system, the nature of the second periodicity is
 controversial. Two different explanations for it have been suggested so far.
 The first one invokes the existence of a strong magnetic field on the surface
 of the WD (Semeniuk et al. 1994, 1995), and the second offers the
 presence of an accretion disk around the primary star (Retter et al. 1995,
 1996).

 \subsection{Superhump interpretation}

 The variation on the 0.085 day is quasi-periodic, as
 we have seen in Fig. 8, and about 4.6 percent longer than the assumed orbital
 period. The periods range and the difference between them are similar to those
 found in SU UMa stars. Table 3 presents the observed periods in this group, as
 well as the corresponding SH relative period excesses (the difference
 between the SH and the binary periods, divided by the binary period),
 taken from Thorstensen et al. (1996).

\begin{table}
\caption{SU Ursae Majoris Periods}
\begin{tabular}{@{}lrrr@{}}

Object & $P_{1}$-orbital & $P_{2}$-superhump & $(P_{2}-P_{1})/P_{1}$ \\
        &       (min)    &       (min)      \\
\\

AL Com    & 81.60 & 82.58 & 0.0120 \\

WZ Sge    & 81.63 & 82.28 & 0.0080 \\

SW UMa    & 81.81 & 83.99 & 0.0266 \\

HV Vir    & 83.51 & 84.46 & 0.0155 \\

WX Cet    & 83.94 & 85.48 & 0.0183 \\

T Leo     & 84.70 & 86.70 & 0.0236 \\

AQ Eri    & 87.75 & 90.25 & 0.0285 \\

V1159 Ori & 89.83 & 92.40 & 0.0286 \\

V436 Cen  & 90.00 & 91.91 & 0.0212 \\

OY Car    & 90.89 & 93.07 & 0.0240 \\

VY Aqr    & 91.41 & 92.70 & 0.0141 \\

ER UMa    & 91.67 & 94.46 & 0.0304 \\

SS UMi    & 97.60 & 101.00& 0.0348 \\

TY Psc    & 98.40 & 101.00& 0.0264 \\

IR Gem    & 98.45 & 102.00& 0.0361 \\

CY UMa    & 100.18& 104.26& 0.0407 \\

FO And    & 103.12& 105.00& 0.0182 \\

HT Cas    & 106.05& 109.55& 0.0330 \\

VW Hyi    & 106.95& 110.49& 0.0331 \\

Z Cha     & 107.28& 111.18& 0.0364 \\

WX Hyi    & 107.73& 111.46& 0.0346 \\

SU UMa    & 109.94& 113.50& 0.0324 \\

V503~Cyg  & 111.90& 116.70& 0.0429 \\

TY PsA    & 121.16& 126.22& 0.0418 \\

YZ Cnc    & 125.00& 131.90& 0.0552 \\

TU Men    & 169.34& 180.83& 0.0679 \\

\end{tabular}
\end{table}

The well known relation between the two periods of SU Ursae Majoris stars
(Stolz \& Schoembs 1981, 1984) is shown in Fig. 9, where the point,
representing the two periods of V1974~Cygni (cross), is also
marked. It is clear that the nova fits well within this relation.

\begin{figure}

\centerline{\epsfxsize=3.5in\epsfbox{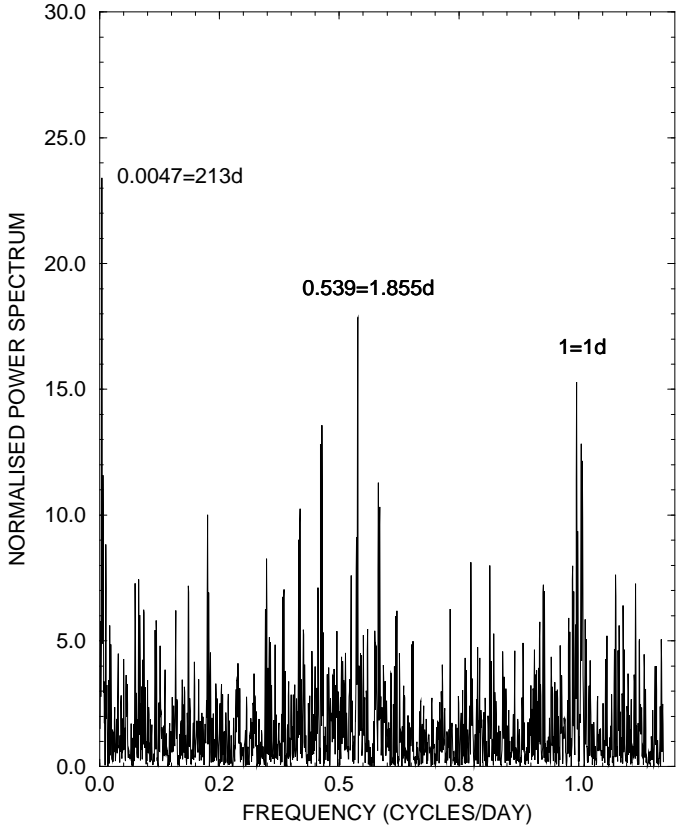}}

 \caption{Superhump relative period excess vs. orbital periods
 in the group of all SU UMa stars for which these values are known. The two
 periods of V1974~Cyg, represented by the cross, obey well this relation.}
\end{figure}

 This fact suggests a possible linkage between the two periods of the classical
 nova V1974~Cygni and the SH phenomenon in SU UMa stars. Further
 indication of the link comes from the 
quasi-periodic nature of the longer periodicity, which is a common 
characteristic of SHs periods (e.g. Patterson et al. 1993a, Leibowitz et al. 
1994). This similarity to the SH phenomenon is quantitative as well. 
In Section 3.5 we estimated the average rate of period change in the LC of 
the nova as $|\dot{P}| \sim 3 \times 10^{-7}$. This value is
 in the range of the values of this parameter as found in permanent SHs:
 $10^{-8} - 5\times 10^{-6}$ (Patterson \& Skillman 1994).

We note here that the amplitude of the longer period ($\sim$ 0.055 mag) is
much lower than superhumps amplitudes in regular SU UMa systems (about
30--40 percent---laDous 1993). It is, however, similar to the typical 
amplitude of the variation in permanent superhumps 
(e.g. Patterson et al. 1993c, Patterson \& Skillman 1994).

 Another two observational features of the LC of V1974~Cyg are also very much
 in line with the SH interpretation. The first one is the third
 periodicity that is found to be present in the 1994 LC (Section 3.3). A third
 period, nearly equal to the binary one but somewhat shorter, has been found in
 a few superhumpers (Patterson et al. 1993c, Harvey et al. 1995, Patterson
 1995 and Patterson et al. 1995a). We do not make an attempt in this work to
 interpret the 1994 third periodicity of Nova~Cyg~92, but the phenomenological
 similarity of it with the "anti-superhump" effect in superhumpers
 may be considered as a further supporting evidence for
 the SH interpretation of the V1974~Cyg LC. The third period found in V1974 
Cyg isn't shorter than the orbital period of the nova as expected from
anti-superhumps, nevertheless it can be interpreted as a real anti-superhump 
if we consider instead its 1 day alias---0.0768 day, which has almost the same 
strength in the PS. The interesting phenomenon of a third periodicity should be 
checked in V1974~Cyg and other superhumpers in the future.

 The second support comes from the apparent displacement towards the blue of
 the high harmonics of the longer periodicity of V1974~Cyg. Again, we do not
 give here an interpretation of this effect. We do draw attention to the
 fact that a similar phenomenon has been observed in PSa of established SU UMa
 systems (Patterson et al. 1995a and Harvey \& Patterson 1995).

\subsection{Magnetic interpretation}

 A different interpretation of the second, longer periodicity in the LC of
 V1974~Cyg has been suggested by Semeniuk et al. (1994, 1995). These authors
 suggest that V1974~Cyg is akin to polar (or intermediate polar) systems. In
 particular they propose that the longer period of N.~Cyg~92 is the spin period
 of the WD in the system. It is manifested as light variation because an
 intense dipole magnetic field on the surface of the WD funnels accreted matter
 onto an accretion column around one of the magnetic poles of the star. The WD
 magnetic dipole is not aligned with the spin axis of the star, and therefore
 the magnetic poles are spinning with the star rotation. The pole onto which
 matter is being accreted is a hot area in the system that changes its aspect
 with the WD spin. Semeniuk et al. speculate that the nova eruption caused the
 WD rotation to get slightly out of synchronization with the orbital frequency,
 hence the difference between the two periodicities. According to this model,
 the tidal effects that synchronize close binary systems are operating now to
 re-synchronize the WD spin to the orbital cycle. The expected observational
 result should be a permanent trend of the longer period to approach the
 shorter value. 

 \subsection{Evaluation of the two interpretations}

 In Section 3.5 we showed that no uniform trend is observed in the difference
 between the two periods of V1974~Cyg along the two years LC of the star.
 Instead, the value of the longer period seems to oscillates
 around a mean value. In particular, in summer 1995 it went against
 the trend expected in the nearly synchronous interpretation, it has {\em
 lengthened} rather than becoming shorter.

 Observationally, similar trend towards synchronization of the second period 
has indeed been detected. In the well studied nova, V1500~Cyg (N.~Cyg~1975) 
and in BY Cam = H0538+608, in which the WD
 is believed to rotate out of synchronization, the shorter of the two periods in
 the LC are indeed becoming longer, evolving towards synchronization with the
 orbital, longer periodicity (Pavlenko \& Pelt 1991, Schmidt \& Stockman 1991,
 Silber et al. 1992, Schmidt, Liebert \& Stockman 1995 and Piirola et al. 1994).

 Further arguments against the a-synchronous interpretation of the second
 period can be raised on the basis of a comparison of other features of the two
 periods phenomenon in V1974~Cyg with the cases where the nearly synchronous
 rotation is commonly believed to be observed. So far there are 3 such cases:
 V1500~Cygni and BY Cam, mentioned before, and RX J19402-1025 
(Patterson et al. 1995b). The binary periods of all 3 systems are around 
3.36 hours (Patterson
 et al. 1995b), whereas that of V1974~Cyg is merely $\sim2$ hours. This
 difference may well be a coincidence, but it is worth being pointed out.

Another significant difference between V1974~Cyg and a-syncronous rotating
polars is that with one exception (RX J19402-1025), all polars have spin
periods shorter than the orbital period (e.g. Patterson 1994). In V1974~Cyg,
on the other hand, the non-orbital period is longer than the orbital one. 
RX J19402-1025 with a negligible period excess 
$(P_{spin} - P_{orbital}) / P_{orbital}\sim 0.3$\%
is not well understood since the accreted matter onto the primary should 
spin up the rotation of the WD.

Additionally, the period excess is quantitatively much larger in 
Nova~Cygni~1992 (about 4.6\%) than in the three a-synchronous cases:
 about -1.8\% in V1500~Cyg (see for example Semeniuk et al. 1977, 
Stockman, Schmidt \& Lamb 1988, Pavlenko 1992), -1.3\% for BY Cam (Silber et 
al. 1992) and 0.3\% in RX J19402-1025 (Patterson et al. 1995b). 

An attempt to explain this difference by the time elapsed since the nova
eruption, assuming all three a-synchronous systems are old novae, is refuted 
by the case of V1500~Cyg. In this nova, the small value has been already 
measured one year after maximum light.

Therefore, it seems to us that the observed data in V1974~Cyg are better
interpreted in terms of disk dynamics, akin to the permanent SH phenomenon, 
rather than reflecting a rotating magnetic field on the surface of the WD in 
the system. We wish to emphasize, however, that our interpretation does not 
exclude the possibility that an intermediate polar model may well be fitted to
this nova system.

 \subsection{The precessing disk model---deriving the binary masses}

 Mineshige, Hirose \& Osaki (1992) applied observational considerations to a
 theoretical equation of Osaki (1985). They present an equation, relating the
 SH periodicity to the parameters of the underlying binary system:

 \begin{equation}
 \bigtriangleup P=\frac{q}{4 \sqrt{1+q} } \eta^{3/2}
 \end{equation}
 where:
 $\bigtriangleup P=(P_{superhump} - P_{orbital}) / P_{orbital}$,
 $q=M_{c}/M_{wd}$
 and $\eta = r_{d} / r_{d,crit}\approx r_{d} / 0.48a = 0.6 - 1$.\\

 It allows to put constrains on the mass ratios of SU UMa stars from the
 observed period excess. Simple approximations lead to the linear relation:

 \begin{equation}
 q\sim(4 - 9) \bigtriangleup P
 \end{equation}
 which describes well Fig. 1 of Molnar \& Kobulnicky (1992).

 Applying equation (2) to Nova~Cyg~1992, with $\bigtriangleup
 P\approx0.0458$, we obtain q = 0.18--0.39. The values from the original
 equation (1) are q = 0.20--0.39. Assuming that the red dwarf fills its Roche
 lobe, we obtain $M_{c}\approx 0.11P(hr)\approx 0.21$ $M_{\odot}$. We can now 
give a final estimation for the WD mass:\\
 0.55--1.07 $M_{\odot}$.

 Paresce et al. (1995) used a combination of 5 theoretical and observational
 considerations to determine the WD mass at the range 0.75--1.1
 $M_{\odot}$. The intersection between this result and ours constraints the WD
 mass to 0.75--1.07 $M_{\odot}$ and q = 0.20--0.28. 

These numbers are inconsistent with the estimation of Austin et al. (1996), 
who concluded that the mass of the WD is about 1.3 $M_{\odot}$ from the 
ejected mass and x-ray considerations.

 \subsection{The superhump period change}

 The O-C diagram presented in Figure 8 and discussed in Section 3.5 indicates
at least  two changes in the $\dot{P}$ value of the longer periodicity of 
V1974~Cyg. The
 lack of observations during the beginning of 1995 prevents the determination of
 the behaviour of the period during this epoch. However, it seems that the
 period oscillates around a mean value.

 Two different cases can be discussed. If the period was fixed at about
 0.084955 day during that time, we might have missed about one cycle, since
 dP/P is about $6\times 10^{-4}$, and the nova was not observed for about
 2000 cycles. In this case the changes in this period may be periodic with a
 period of about 7500 cycles or 21 months.

 Alternatively, during the observational gap the period may have changed its 
sign twice,
 increasing first, and then decreasing again. Some support for this notion
 comes from the first lonely point in the O-C diagram, which suggests another 
change in $\dot{P}$. However, this point may
 be also one cycle misplaced. The PS of all O-C points is presented
 in Fig. 10. Its peak is found at 213 day or about 2500 cycles. This is
 the result of the later hypothesis for the period changes.

 \begin{figure}

 \centerline{\epsfxsize=3.5in\epsfbox{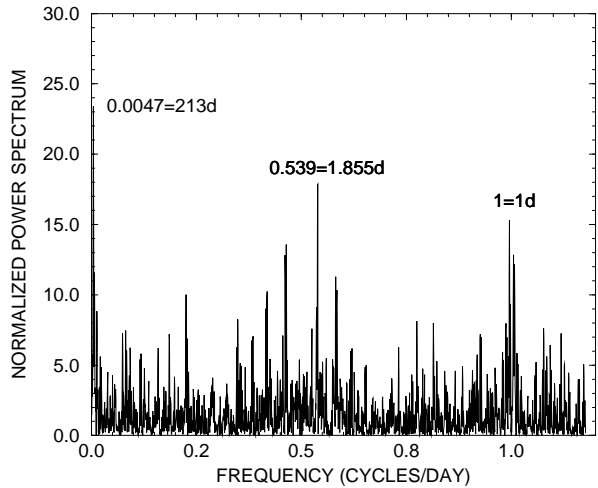}}

 \caption{The power spectrum of O-C points of the longer period. At the left
 side of the graph there is the probable periodicity of about 213 days which
 reflects the long term changes in the 0.0849767 day period. The beat period
 between the two short-term periods (1.855 day) is also seen at the center.}
 \end{figure}

 If one of these two scenarios is true, $\dot{P}$ of the quasi-period may be
 periodic, and we expect the 0.0849767 day period to decrease around July,
 1996. We emphasize that, this prediction is valid for the two cases. The two
 possibilities will become distinguishable towards the end of 1996, when an
 increase of the period will indicate, that the second possibility is right,
 while a constant period will be consistent with a longer periodicity of
 $\dot{P}$ as the first case.

 This prediction should be taken with caution, because we interpret the
 0.0849767 day period as a SH period. Changes in SHs are very
 common in SU UMa stars, and in permanent superhumpers. However, in the few 
well observed cases
 known so far, these changes seem to be in the nature of a drift in the period
 rather than being truly periodical. (See for example the O-C diagrams of AM
 CVn in Patterson et al. 1993c).

 Another interesting feature of the PS of the O-C diagram 
is the strong peak around the periodicity of 1.855 day, which corresponds 
to the beat period between the two periods of the nova, the binary and 
the longer one. This fact is an additional confirmation for the presence 
of the two independent frequencies, and it can explain the scatter of 
the points in Fig. 8.

 \subsection{The overall picture and the orbital dip}

 The development of the optical photometric features of V1974~Cyg can be simply
 explained by a model of two distinct major light sources in the system. During
 the observations in 1993 the variation in the light of the nova was due to
 reflection from the secondary star in the system, illuminated by the intense
 radiation from the WD vicinity (DeYoung \& Scmidt 1994). As the
 radiation intensity of the primary weakened, with the decay of the erupting
 nova, another periodicity became dominant in the LC of the system in
 1994.

 There is also a significant change in the structure of the binary cycle
 itself. To the rather smooth, roughly sinusoidal variation in 1993, a shallow
 eclipse like feature has been added in 1994 and 1995 to the binary LC, half a
 cycle after minimum light. 

According to our suggestion that the second periodicity is a superhump-like
phenomenon, the light source that appeared in 1994, is a precessing accretion
disk. This would mean that the classical nova V1974~Cyg entered a state of 
permanent SH (Osaki 1996) a few months after its eruption.

How can the structure of the photometric binary cycle be understood within 
this framework?
We think that the orbital dip at phase 0.5 may give us a clue. Such orbital 
dips are common in soft x-ray sources, where they appear sometimes not only 
in an average over many cycles, but even at at individual light curves 
(Callanan, Grindlay \& Cool 1995, Kato et al. 1995, Bailyn 1992). 
Callanan et al. interpreted the modulation in 4U 1916-05 as the obscuration 
of the accretion disk by the bright spot. This interpretation may also be 
valid for the variation in the LC of V1974~Cyg.

\section{Summary and Conclusions}

The similarity between Nova~Cygni~1992 and systems showing permanent 
superhumps is expressed by the following facts:\\
1. The presence of  two independent periodicities in the light curve, which
lie in the short-term range of SU UMa periods.\\
2. The difference between the two periods is a characteristic of
SU UMa systems, in which the longer period is related to the superhump
variation. The two periods obey the relation of Stolz \& Schoembs (1981, 1984) 
for the two periods of SU UMa systems\\
3. The amplitude in the longer period is similar to the values found in 
permanents superhumps.\\
4. The assumed superhump variation is in fact quasi-periodic with $|\dot{P}|$ 
typical to permanent superhumps.\\
5. The probable existence of a third periodicity in the power spectrum of 
the observations in 1994, which may be interpreted as an anti-superhump 
period.\\
6. The higher harmonics of the longer period are slightly misplaced towards
the blue. Similar shift is found in superhumpers.

The explanation of the longer period as caused by the rotation of the 
magnetic white dwarf is inconsistent with the fact that the system does
not march towards synchronization. Essentially only one polar system with
$P_{spin} > P_{orbital}$ is known (RX J19402-1025), but with negligible 
period excess (0.3\%).
 
The features described above may fit the intermediate polar model, if
another short periodicity---the spin period may be found in the future.
 
Our results suggest, therefore, the presence of an accretion disk in the
V1974~Cyg stellar system no later than 30 months after the outburst of the nova.
The system is in a state of superoutburst (permanent superhump) for more than
two years, up to the present time.
 
The precessing disk model and previous results constrain the binary
parameters to the domains:\\
0.75 $\leq M_{wd}/M_{\odot} \leq 1.07$ and $0.20 \leq q \leq 0.28$.

\section{Acknoledgements}

We thank John Dan and the Wise Observatory staff for their expert assistance 
with the observations.We are greatful to James DeYoung and Irena Semeniuk 
for sending us their data.
We would also like to thank Tsevi Mazeh for helpful remarks
and Domitilla De-Martino for discussing magnetic field properties.
Special thanks are due to Shai Kaspi for general help fighting the computers.
This research has made use of the AFOEV database, operated at CDS, France.

Astronomy at the Wise Observatory is supported by grants from the Israeli 
Academy of science.


\begin{thebibliography}{99}

 \bibitem{b9} Annuk K., Kolka I., Leedjarv L., 1993, A\&A, 269, L5.

 \bibitem{b9} Austin S.J., Wagner R.M., Starrfield S., Shore S.N., 
Sonneborn G., Bertram R., 1996, Astro. J. 111/2, 869.

 \bibitem{b2} Bailyn C.D., 1992, Ap.J., 391, 298.

\bibitem{b2} Callanan P.J., Grindlay J.E., Cool A. M., 1995, PASJ, 47, 153.

 \bibitem{b2} Charles P.A., Kidger M.R., Pavlenko E.P. Prokofiera, V.V.,
  Callanan P.J., 1991, M.R.A.S., 249, 567.

\bibitem{b9} Chochol D., Hric L., Urban Z., Komzik R., Grygar J., Papousek
J., 1993, A\&A, 277, 103.

 \bibitem{b12} Collins P., 1992, IAUC, 5454.

 \bibitem{b12} DeYoung J.A., Schmidt R.E., 1993, IAUC, 5880.

 \bibitem{b2} DeYoung J.A., Schmidt R.E., 1994, Ap.J., 431, L47.

  \bibitem{b11} laDous C., in Cataclysmic Variables \& Related Objects, 
Hack M., laDous C. (ed.), 1993, Centre National de la Recherce Scientifique, 
Paris, France.

 \bibitem{b3} Harvey D.A., Patterson J., 1995, PASP, 107, 1055.

 \bibitem{b3} Harvey D.A., Skillman D.R., Patterson J., Ringwald F.A.,
 1995, PASP, 107, 551.

\bibitem{b12} Hurst G.M., 1992, IAUC, 5526.

\bibitem{b1} Kaspi S., Ibbetson P. A., Mashal E., Brosch N. 1995, 
Wise Obs. Tech. Rep., No 6.

\bibitem{b3} Kato T., Mineshige S., Hirata R., 1995,  PASP, 47, 31.

\bibitem{b12} Kidger M., 1992, IAUC, 5526.

 \bibitem{b2} Leibowitz E.M., Mendelson H., Bruch A., Duerbeck H.W., 
Scitter W.C., Richter G.A., 1994, Ap.J., 421, 771.

 \bibitem{b3} Mineshige S., Hirose M., Osaki Y., 1992,  PASJ, 44, L15.

\bibitem{b3} Molnar L.A., Kobulnicky H.A., 1992, Ap.J., 392, 678.

\bibitem{b4} Netzer H. et al., 1996, MNRAS, 279, 429.

\bibitem{b9} Osaki Y., 1985, A\&A, 144, 369.

\bibitem{b3} Osaki Y., 1996,  PASP, 108, 39.

 \bibitem{b9} Paresce F., Livio M., Hack W., Korista K., 1995, A\&A,
 299, 823.

\bibitem{b3} Patterson J., 1994, PASP, 106, 209.

\bibitem{b3} Patterson J., 1995, PASP, 107, 657.

 \bibitem{b3} Patterson J., Richman H., 1991, PASP, 103, 735.

 \bibitem{b3} Patterson J., Skillman D.R., 1994, PASP, 106, 1141.

 \bibitem{b3} Patterson J. Bond H.E., Grauer A.D., Shafter A.W.,
 Mattei J., 1993a, PASP, 105, 69.

\bibitem{b3} Patterson J., Halpern J., Shambrook A., 1993b, Ap.J., 419, 803.

 \bibitem{b3} Patterson J. Thomas G.R., Skillman D., Diaz M.,
 Suleimanov V.F., 1993c, Ap.J.Suppl., 86, 235.

 \bibitem{b3} Patterson J, Jablonski F., Koen C., O'Donghue D.,
 Skillman D.R., 1995a, PASP, 107, 1183.

 \bibitem{b3} Patterson J., Skillman D.R., Thorstensen J.R.,
 Hellier C., 1995b, PASP, 107, 307.

 \bibitem{b3} Pavlenko E.P., 1992, Izv. Krym. Astrofiz., Obs., 86, 55.

 \bibitem{b3} Pavlenko E.P., Pelt J., 1991, Astrophizika, 34/2, 169.

 \bibitem{b3} Piirola V., Cogne G.V., Takalo L., Larsson S.,
 Vilhu O., 1994, A\&A, 283/1, 163.

\bibitem{b12} Retter A., Ofek E.O., Leibowitz E.M., 1995, IAUC, 6158.

\bibitem{b12} Retter A., Leibowitz E.M., Ofek E.O., 1996, 
Proceedings of the 158th Colloquium of the IAU, held at Keele, 
Evans N. and Wood J. (eds.), Kluer Academic Publishers.

 \bibitem{b6} Scargle J.D., 1982, Ap.J., 263, 835.

 \bibitem{b10} Schmidt G.D., Stockman H.S., 1991, Ap.J., 371, 749.

 \bibitem{b10} Schmidt G.D., Liebert J., Stockman H.S., 1995, Ap.J., 441, 414.

 \bibitem{b10} Semeniuk I., Kruszewski A., Schwarzenberg-Czerny, A.,
 Chlebowski T., Mikolajewski M., Wolczk J., 1977, Acta Astron., 27/4, 301.

 \bibitem{b10} Semeniuk I., Pych W., Olech A., Ruszkowski M., 1994,
 Acta Astron., 44, 277.

 \bibitem{b10} Semeniuk I., DeYoung J.A., Pych W., Olech A.,
 Ruszkowski M., Schmidt R.E., 1995, Acta Astron., 45, 365.

 \bibitem{b3} Skillman D.R., Patterson J., Thorstensen J.R., 1995, PASP, 
107, 545.

 \bibitem{b3} Silber A., Bradt H.V., Ishida M., Ohashi T., Remillard R.A.,
1992, Ap.J., 389, 704.

\bibitem{b3} Solheim J.E. et al., 1996, submitted to A\&A.

\bibitem{b3} Stetson P.B., 1987, PASP, 99, 191.

\bibitem{b3} Stockman H.S., Schmidt G.D., Lamb D.Q., 1988, Ap.J., 332, 282.

 \bibitem{b8} Stolz B., Schoembs R., 1981, IBVS, 2029.

 \bibitem{b9} Stolz B., Schoembs R., 1984, A\&A, 132, 187.

 \bibitem{b3} Thomas G.R., 1993, B.A.A.S., 25, 909.

 \bibitem{b3} Thorstensen J.R., Patterson J., Shambrook A.,
 Thomas G.R., 1996, PASP, 108, 73.

 \bibitem{b3} White J.C.2, Honeycutt R.K., 1992, B.A.A.S., 24, 1284.

 \bibitem{b3} White J.C.2, Honeycutt R.K., Horne K., 1993, Ap.J., 412, 278.

 \bibitem{b3} White N.E., 1989, A\&A Rev., 1, 85.

\bibitem{b4} Whitehurst R., 1988, MNRAS, 232, 35.

 \bibitem{b3} Zhang Z.Y., Chen J.S., 1992, A\&A, 266, L9.

\end{thebibliography}
\end{document}